\input epsf
\documentstyle[preprint,aps,eqsecnum,tighten]{revtex}

\def\be{\nopagebreak[3]\begin{equation}}
\def\ee{\end{equation}}
\def\ba{\nopagebreak[3]\begin{eqnarray}}
\def\ea{\end{eqnarray}}
\def\nl{\nonumber \\}

\def\a{\alpha} 
 
\def\g{\gamma} 
\def\e{\epsilon} 
 
\def\G{\Gamma} 
\def\S{\Sigma}
\def\Com{\bf C} 
 
\def\x{\vec x}
\def\y{\vec y}
\def\s{{\vec s}}

\def\Ethree{{}^3\!E[f]}

\def\Etwo{{}^2\!E[S,f]}
\def\Etwog{{}^2\!E[S,g]}

\def\Eh{\hat{E}^a_i}
\def\Xthree{{}^3\!X [f]}
\newcommand{\teta}{\rlap{\lower2ex\hbox{$\,\tilde{}$}}\eta{}}

\newcommand{\tE}{E}

\def\d{{\rm d}}
\def\Ta{{\rm T}_{\alpha}}

\def\tr{{\rm Tr}}
\def\A{{\cal A}}
\def\Ab{\bar{\A}}
\def\H{{\cal H}}
\def\C{{\cal C}}
\def\cyl{{\rm Cyl}}
\def\cylg{{\rm Cyl}_\gamma}

\def\F{{\cal F}}
\def\V{{\cal V}}

\begin{document}
\title{Quantum Theory of Geometry III: \\
Non-commutativity of Riemannian Structures}
\author{Abhay Ashtekar${}^{1,3}$\thanks{Electronic address: 
ashtekar@phys.psu.edu},
Alejandro Corichi${}^{1,2}$\thanks{Electronic address: 
corichi@nuclecu.unam.mx},
and
Jos\'e A. Zapata${}^{1}$\thanks{Electronic address:
zapata@phys.psu.edu}
}

\address{$1$ Center for Gravitational Physics and Geometry\\
Department of Physics, Penn State,\\ 
University Park, PA 16802-6300, USA
}
 
\address{$2$ Instituto de Ciencias Nucleares\\ Universidad Nacional
Aut\'onoma de M\'exico\\ A. Postal 70-543, M\'exico D.F. 04510,
M\'exico }

\address{$3$ Erwin Schr\"odinger International Institute 
for Mathematical Physics\\
Boltzmanngasse 9, A-1090 Vienna, Austria
}
\maketitle

\begin{abstract}

The basic framework for a systematic construction of a quantum theory
of Riemannian geometry was introduced recently. The quantum versions
of Riemannian structures --such as triad and area operators-- exhibit
a non-commutativity. At first sight, this feature is surprising
because it implies that the framework does not admit a triad
representation. To better understand this property and to reconcile it
with intuition, we analyze its origin in detail. In particular, a
careful study of the underlying phase space is made and the feature is
traced back to the classical theory; there is no anomaly associated
with quantization. We also indicate why the uncertainties associated
with this non-commutativity become negligible in the semi-classical
regime.

\end{abstract} 

\section{Introduction}
\label{sec1}

This article is a continuation of \cite{1,2} which will be referred to
as papers I and II respectively. Unless otherwise specified, we will
use the same notation and conventions.

Let us begin with a brief summary. In a non-perturbative treatment of
quantum gravity, a preferred classical metric is not
available. Therefore, one has to develop the appropriate field theory
without any reference to a background geometry. One possibility is to
use the canonical approach based on connections. Here, the
configuration variable is an $SU(2)$ connection $A_a^i(\x)$ on a
three-manifold $\S$ (which serves as the kinematic arena). The
momentum variable is a triad ${E}^a_i(\x)$ with density weight one.
(Indices $a,b,c,...$ refer to the tangent space of $\S$ and
$i,j,k,...$ to the $su(2)$ Lie-algebra.) Since the classical
configuration space $\A$ is the space of smooth connections on $\S$,
the quantum configuration space $\Ab$ turns out to be a space of
suitably generalized connections on $\S$. To obtain the Hilbert space
$\H$ of quantum states and geometric operators thereon, one needs a
functional calculus on $\Ab$ which also does not refer to a background
metric (or any other field).

The necessary tools were developed in a series of papers by a number
of authors \cite{3,4,5,6,7,8,9,10,rs,b,hall,alm2t}. (Much of the
motivation for this work came from the `loop representation'
introduced earlier by Rovelli and Smolin \cite{11}.)  It turns out
that $\Ab$ admits a natural diffeomorphism invariant measure
$\mu_{\circ}$ \cite{4,5,6,7,9} and the Hilbert space $\H$ can be taken to
be the space $L^2(\Ab, \d\mu_{\circ})$ of square-integrable functions
on $\Ab$. Physically, $\H$ represents the space of kinematic quantum
states, i.e., the quantum analog of the full phase space.  Using the
well-developed differential geometry on $\Ab$ \cite{8}, one can then
define physically interesting operators on $\H$. In particular, one
can introduce, in a systematic manner, operator-valued distributions
$\Eh$ corresponding to the triads \cite{1}. As in classical
differential geometry, these are the basic objects of quantum
Riemannian geometry. Specifically, operators corresponding to area,
volume and length are constructed by regularizing the appropriate
products of these triad operators \cite{1,2,12}. (For related
frameworks, see \cite{13,14,15,16}).

Being density weighted, the triads ${E}^a_i$ are duals of pseudo
two-forms $e_{abi}:= \eta_{abc}{E}^c_i$.  In the quantum theory,
therefore, one might expect that they should be smeared against
Lie-algebra-valued test fields $f_i$ with support on {\it
two-dimensional surfaces}. This expectation turns out to be correct%
\footnote{In this approach to quantum geometry, there is a remarkable
synergy between geometry and analysis: in the regularization
procedure, well-defined operators result when n-forms are integrated
on n-manifolds. Thus, the operators that code information about
connections are holonomies $\hat{h}[\alpha]$, obtained by integrating
the connection one-forms along one dimensional curves. The triad
two-forms are naturally regulated through a two dimensional
smearing. This feature is deeply connected with the underlying
diffeomorphism invariance of the theory. By contrast, in the quantum
theory of Maxwell fields in Minkowski space-time, for example, using
the geometrical structures made available by the background metric,
one uses a three dimensional smearing for both connection
one-forms and electric field two-forms.}
\cite{1}. However, somewhat surprisingly, the resulting operators 
${}^2\!\hat{E}[S,f]$ turn out not to commute. For example, for
operators smeared by two different test fields on the {\it same}
two-surface, we have:
\be \label{1}
[{}^2\!\hat{E}[S, f] , {}^2\!\hat{E}[S, g] \not= 0
\ee
if the Lie-bracket $[f,g]^i = {\e^{i}}_{jk}f^jg^k$ fails to vanish.
If the two operators are smeared along two distinct surfaces, the
commutator is again non-zero if the two surfaces intersect and the
Lie-bracket of the corresponding test fields is non-zero on the
intersection.

This feature is at first surprising --even disturbing-- because it
implies that we can not simultaneously diagonalize all the triad
operators. Could it be related to the fact that triads are only
covariant rather than invariant under $SU(2)$ rotations? Would this
non-commutativity disappear if one dealt only with manifestly gauge
invariant objects? The answer is in the negative: the
non-commutativity extends also to, e.g., the area operators which are
gauge-invariant. In particular, if we have two surfaces $S_1$ and
$S_2$ which intersect along a line, the corresponding area operators
$\hat{A}[S_1]$ and $\hat{A}[S_2]$ fail to commute. The commutator
$[\hat{A}[S_1], \hat{A}[S_2]]$ is non-zero only on those states which
(in the terminology of paper I) have a four or higher valent vertex on
the intersection. It is true that, heuristically, such states are
`non-generic'. Nonetheless, they constitute an infinite dimensional
subspace of the space of gauge invariant states. Hence, even if we
restrict ourselves to the gauge invariant context, the
non-commutativity persists and makes it impossible to simultaneously
diagonalize all the geometric operators. Thus, {\it the quantum
Riemannian geometry that arises in this framework is genuinely
non-commutative} and, at a fundamental level, one must face the
Heisenberg uncertainties associated with geometric
quantities. However, because the commutators fail to vanish only on
certain rather special states, as one might expect, the quantum
uncertainties turn out to be completely negligible in the
semi-classical regime.

Nonetheless, the fact that the triad --and hence the metric--
representation fails to exist in this approach is striking and was not
fully appreciated in the early literature on the subject.  It is
important to understand its origin. Does this feature arise because of
some subtleties associated with the classical Poisson brackets? Or, is
this a quantum anomaly? In either case, what precisely are the
underlying assumptions that lead to this non-existence of the triad
representation? The purpose of this paper is to address these issues
in a systematic fashion. We should emphasize that the consistency of
the quantum theory is not in question here. The construction of the
Hilbert space and the introduction of the operators has been carried
out in a rigorous fashion. The regularization procedure is natural and
tight. Rather, one wishes to reconcile the results of that analysis
with one's intuition, particularly with what one knows about the phase
space structure of the classical theory.

The paper is organized as follows.  In section \ref{sec2}, we will
discuss certain subtleties associated with phase space structures: the
`obvious' choice of configuration and momentum variables turns out to
be inappropriate in view of gauge and diffeomorphism invariance and,
furthermore, the `obvious' choice of Poisson brackets between the
appropriate variables leads to inconsistencies. One must therefore
find an appropriate substitute of the naive Poisson brackets. The
required Lie algebra is presented in section \ref{sec3}.  We will find
that the Lie-bracket structure of the classical theory simply mirrors
that found in paper I for the quantum theory. Thus, there is {\it no}
quantum anomaly. The main results are then examined from several
angles to reconcile them with intuitive expectations.  Section
\ref{sec4} contains a summary and remarks. In particular, we
elucidate why the non-commutativity of area operators does not lead to
unwarranted uncertainties in the semi-classical regime.

\section{Phase space structures: Subtleties}
\label{sec2}

In section \ref{sec2.1}, we recall the structure of the classical
phase space and fix notation. In section \ref{sec2.2}, we point out
that the underlying diffeomorphism invariance now leads us to
phase space variables which are rather different from those
used, e.g., in the Maxwell theory in Minkowski space: now the
configuration variables are Wilson loops and the momenta are triads
smeared in {\it two} dimensions. From the viewpoint of standard phase
space discussions, these functionals are `singular' as they are
obtained by smearing the basic canonical variables in one or two
dimensions, rather than three.  Consequently, one's naive expectation
on the structure of their Poisson algebra may be incorrect. In
\ref{sec2.3}, we will see that this is indeed the case: the naive
Poisson brackets fail to satisfy the Jacobi identity.

\subsection{Phase space}
\label{sec2.1}

Fix an oriented, analytic%
\footnote{The assumption of analyticity is made only because it will
simplify certain technicalities in sections \ref{sec2} and 
\ref{sec3}. We believe that the entire discussion can be carried
over to the smooth and piecewise linear categories by appropriate
modifications along the lines of \cite{9,17} respectively.}
three-manifold $\S$. We will assume that $\S$ is either compact or
that the various fields satisfy suitable boundary conditions at
infinity, the details of which will play no role in our
analysis. Because $\S$ is three-dimensional and oriented, the
principal $SU(2)$ bundle over $\S$ is trivial. Therefore, we can
represent $SU(2)$ connections on the bundle by $su(2)$-valued
one-forms $A_{aC}{}^D$, where $a$ is the one-form index and $C,D$
refer to the fundamental representation of $SU(2)$.  For notational
simplicity, we will often set ${A_{aC}}^D = A_a^i\tau_{iC}{}^D$, where
the anti-Hermitian $\tau_i$ are related to the Pauli matrices
$\sigma_i$ via $2i\tau_i = \sigma_i$. (Thus, $-2
\tr\tau_i\tau^j = \delta_i^j$ and $[\tau_i, \tau_j] = \e_{ijk}
\tau^k$.)  We will assume that all fields on $\S$ are smooth.

The configuration space $\A$ consists of all smooth connections
$A_a^i$ on $\S$ satisfying the boundary conditions. Thus, $\A$ is
naturally an affine space. The phase space is the cotangent bundle
over $\A$.  The momenta are represented by smooth vector densities
${E}^a_i$ of weight one on $\S$, or equivalently, by the triplet of
two-forms $e_{ab}^i= \eta_{abc} {E}^{ci}$.  The fundamental Poisson
brackets are given by:
\ba \label{2}
\{A_a^i(\x), A_b^j(\y)\} = 0\quad &;& \quad 
\{ {E}^a_i(\x), {E}^b_j(\y)\} =0; \nl
\{A_a^i(\x), {E}^b_j(\y)\} &=& G\, \delta_a^b\, \delta^i_j\,
\delta^3(\x,\y), 
\ea
where $\x$ and $\y$ denote points on $\S$ and $G=8\pi\,G_{\rm N}$,
where $G_{\rm N}$ is Newton's constant. In particular, the triads
Poisson-commute among themselves.  This is why it is at first very
surprising that the quantum triads fail to commute. The naive
conclusion would be that the specific quantization is anomalous;
vanishing Poisson brackets in (\ref{2}) go over to non-trivial
commutators. However, we will see in sections \ref{sec2.2} and
\ref{sec2.3} that this is not the case.

For now, let us only note the meaning of the Poisson brackets. The
fields $A_a^i$ and $\tE^a_i$ on $\S$ are {\it not} functions on the
phase space.  Therefore, the Poisson brackets (\ref{2}) between them
can only be interpreted in the distributional sense. That is, given
any vector density ${v}^a_i$ which takes values in the dual of the
$su(2)$ Lie algebra, and a one-form $f_a^i$ which takes values in the
$su(2)$ Lie algebra, we can naturally define smooth functions
\be \label{3} 
{}^3\!A[v]\, := \int_\S \d^3\!x\, A_a^i(\x) {v}^a_i(\x),\quad\quad
{}^3\!E[f]\, := \int_\S \d^3\!x \, {E}^a_i(\x) f_a^i(\x) 
\ee
on the phase space, where the superscript $3$ makes it explicit that
we have used three dimensional test fields to smear the basic
variables. Equation (\ref{2}) is then just a short-form for the
following Poisson brackets between these well-defined configuration
and momentum {\it functions}:
\ba\label{4}
\{ {}^3\!A[v], {}^3\!A[v'] \} = 0\quad &;& 
\quad \{ {}^3\!E[f], {}^3\!E[f'] \} = 0;\nl
\{ {}^3\!A[v], {}^3\!E[f] \} &=& G \int_\S \d^3\!x\, {v}^a_i
f_a^i\, .
\ea
A priori, relations (\ref{2}) do not say anything about the Poisson
brackets between `singular' functions obtained by integrating $A_a^i$
and $\tE^a_i$ by distributional smearing fields.

In linear field theories in Minkowski space, there is no need to
consider `singular' smearings. Indeed, one generally begins with the
Abelian algebra generated by finite complex linear combinations of
finite products of configuration variables ${}^3\!A[v]$. These are
generally referred to as cylindrical functions on the configuration
space $\A$. Because of the underlying linear structure of $\A$, one
can apply the standard Kolmogorov \cite{18} theory to integrate these
functions. Thus, to construct the quantum theory, one has to select a
suitable cylindrical measure, define an Hermitian inner product
between cylindrical functions and Cauchy complete the space to obtain
the quantum Hilbert space $\H$. On this Hilbert space, the
configuration operators ${}^3\!\hat{A}[v]$ act naturally by
multiplication. Finally, to complete the kinematic set-up, one defines
the action of the momentum operators ${}^3\!\hat{E}[f]$ so that the
Poisson brackets (\ref{4}) are taken over to $i\hbar$ times the
commutators. (Typically, this action is obtained by studying the
action of the Hamiltonian vector fields generated by ${}^3\!E[f]$ on
cylindrical functions.)

In the present case, however, the kinematic symmetries of the theory
--the $SU(2)$ gauge invariance and the $\S$-diffeomorphism
invariance-- make this strategy unsuitable. (See, e.g., \cite{19} for
details.)  More precisely, the simple configuration variables in
(\ref{3}) fail to be gauge covariant and are therefore unsuitable in
the non-Abelian context (unless one manages to fix the gauge 
completely). Furthermore, there is also a problem with respect to the
diffeomorphism invariance: the above cylindrical functions fail to be
integrable with respect to any of the known diffeomorphism invariant
(generalized) measures.  Hence, the simple Poisson algebra (\ref{4})
is no longer suitable as the starting point for quantization. One has
to find an appropriate substitute.

\subsection{Appropriate phase space variables}
\label{sec2.2}

The strategy \cite{20,11,3} that has been most successful is to
construct the configuration observables through holonomies of
connections. Perhaps the simplest possibility is to use, in place of
${}^3\!A[v]$, the Wilson loop functionals, i.e. traces of holonomies
of connections around closed, piecewise analytic loops, $T[\a]$:
\be\label{5}
T[\a] := \frac{1}{2}\, \tr\,\, {\cal P}\,\exp\,(-\oint_\a A_a \d s^a)
\ee
where ${\cal P}$ stands for `path ordered'. The algebra generated by
these functions is called the holonomy algebra \cite{3}.

While this construction of the algebra is simple and direct, in order
to do functional analysis on this space of functions of connections,
one has to introduce, in addition, considerable technical
machinery. This involves the introduction of the appropriate notion
of `independent loops' and techniques for decomposing arbitrary
piecewise analytic loops into a finite number of independent ones. (For
details, see \cite{4}).  Furthermore, by construction, all elements of
the holonomy algebra are gauge invariant. Since we wish to examine the
action of canonical transformations generated by triad functions
--which are only gauge covariant-- we need a larger arena to work
with. We will therefore proceeds as follows \cite{5,8}.

Denote by $\g$ a graph in $\S$ with $N$ analytic edges. (For our
purposes, an intuitive understanding of these notions will
suffice. For precise definitions, see, e.g., \cite{1,8}.) Let us fix a
global cross-section of our principal $SU(2)$ bundle. Then, every
connection $A$ in $\A$ associates with each edge $e$ of $\g$ an
element $h_e(A)$ of $SU(2)$, its holonomy along the edge
$e$. Therefore, given any complex-valued, smooth function $c$, 
on $[SU(2)]^{N}$,
$$ c:\,\, [SU(2)]^{N}\, \longrightarrow\,  \Com $$
we acquire a function $C_{\g}$ on $\A$:   
\be\label{6}
C_\g (A)\, := \, c(g_1, \,...\, g_{N})
\ee
(Since the function on $\A$ depends not only on $\g$ but also on our
choice of $c$, strictly, it should be denoted as $C_{\g,c}$. However,
For notational simplicity, we will only retain the suffix $\g$.)  These
configuration variables capture only `finite dimensional pieces' from
the infinite dimensional information in the connection field $A_a^i$;
they are sensitive only to what the connection does on the edges of
the graph $\g$. Following the terminology used in linear field
theories, they are called {\it cylindrical functions}. The space of
cylindrical functions associated with any graph $\g$ is denoted by
$\cylg$. The Wilson loops functionals $T_\a$ associated with closed
loops $\a$ that lie entirely in $\g$ clearly belong to $\cylg$. (By
suitably restricting the form of functions $c$, we can make $\cylg$
the algebra generated by these Wilson loop functionals. However, we
shall not require this.)  The space $\cylg$ is quite `small': it only
contains the configuration observables associated with the `lattice
gauge theory' defined by the graph $\g$. However, as we vary $\g$,
allowing it to be an arbitrary graph (with a finite number of analytic
edges), we obtain more and more configuration variables. Denote the
union of all these by Cyl. This is a very large space. In particular,
it suffices to separate points of $\A$. {\it We will take {\rm Cyl} to
be the space of our configuration variables.}

Next, we wish to consider the triads. Let us begin with
$\Ethree$. What are the Poisson brackets between $\Ethree$ and
functions in Cyl? Since the Hamiltonian vector field of $\Ethree$ is
well-defined, the calculation is easy to perform.  One obtains:
\ba\label{7} 
\{ C_\g, \, \Ethree \} 
&=& G \int_\S \d^3\!x\, f_a^i(\x) \frac{\delta
C_\g}{\delta A_a^i(\x)}\nl 
&=& G \sum_{I=1}^{N}\, \int_{e_I} \d t_I \, \dot{e}^a_I(t_I) 
f_a^i(e_I(t_I))\, [h(1,t_I)\tau_i h(t_I,0)]^A_B\,\,
\frac{\partial{c}}{\partial{h_{I}}{}^A_B}\, , 
\ea
where $t_I$ is a parameter along the edge $e_I$ that runs between $0$
and $1$, and $h(t_I, 0)$ is the holonomy along $e_I$ from $t_I =0$ to
$t_I =1$. Thus, as expected, the Poisson bracket is
well-defined. However, unfortunately, the result is {\it not} a
cylindrical function: because of the $t_I$ integration involved, the
right side requires the knowledge not just of a finite number of
holonomies, $h_1,\, ...\, ,h_{N}$, but of $N$-continuous parameter
worth of them, $h_I(t)$.

Recall from section \ref{sec2.1} that a standard quantization strategy
is to construct the Hilbert space of states from configuration
variables. In the resulting quantum theory, the action of the momentum
operators is then dictated by the action, on configuration variables,
of the Hamiltonian vector fields generated by momenta. Therefore, it
is highly desirable that the configuration variables be closed under
this action.  Unfortunately, this is not possible with $\Ethree$ as
our momenta.

Note, however, that a drastic simplification occurs if the smearing
field $f_a^i$ is distributional with two-dimensional support. Let us,
for definiteness introduce local coordinates on $\S$ and assume that
the smearing field has the form:
$${}^{\e}\!f_a^i(\x) =  h_{\e}(z) (\nabla_a z) f^i(x,y), $$
where $h_\e(z) $ tends to $\delta(z)$ in the limit that the parameter
$\e$ goes to zero. Then, in the limit as $\e$ goes to zero, the
continuous integral over $t_I$ in (\ref{7}) collapses to a finite sum
over the points $p$ at which $e_I$ intersects the surface $z=0$ and
the right side is again a cylindrical function. 

Let us evaluate this limit explicitly.  At each of the intersection
points $p$, one has a number of edges, $e_{I_p}$, with $I_p = 1,2,\,
...\, ,n_p$. If all these edges are oriented away from the surface, we
have:
\be\label{10}
\lim_{\e\mapsto 0}\{ C_\g, \, {}^3\!E[{}^\e\!f] \} 
= \frac{G}{2} \sum_{p}\, \sum_{I_p}\, \kappa({I_p})\, f^i(p)
[h_{I_p}\, \tau_i]^A_B\,\, 
\frac{\partial{c}}{\partial{h_{I_p}}{}^A_B}\, , 
\ee
where the constant $k({I_p})$ equals $+1$ if the edge lies entirely
above $S$, $-1$ if it lies below $S$ and $0$ if it is tangential to
$S$.  If an edge is oriented towards the surface, the only change is
that the factor $[h_{I_P}\tau_i]$ is replaced by $[\tau_i h_{I_P}]$.
Finally, note that 
$$ [h_{I_p}\, \tau_i]^A_B\,\, 
\frac{\partial{c}}{\partial{h_{I_p}}{}^A_B}\, $$
is just the result of the action of the $ith$ left invariant vector
field on the $I_p$th argument of the function $c(h_1,\, ...\, ,
h_{I_p},\, ...\,)$ on $[SU(2)]^{N}$. Similarly, if an edge $I_p$ is
oriented towards the surface, we get the action of the $ith$ right
invariant vector field. The details of this argument can be found in
section 3.1 of \cite{1}. Here, we only note that there are no
contributions from edges which lie in the limiting surface $z=
0$. For, if an edge $e_I$ is tangential, we have $\dot{e}^a_I
\nabla_a{z} = 0$ before taking the limit $\e \mapsto 0$. Hence these
edges make no contribution; $\kappa(I_p)$ vanishes in this case.

Thus, if we use distributional test-fields with two dimensional
support and evaluate the Poisson bracket of the resulting smeared
triad with cylindrical functions via a limiting procedure, we find
that the Poisson bracket closes on $\cyl$. Hence, it is natural to use
two-dimensionally smeared triads
\be \label{e2}
\Etwo := \int_S e_{abi}f^i \d S^{ab}
\ee
as the momentum variables. However, the limiting procedure used to
pass from three dimensionally smeared triads to the two dimensionally
smeared ones is technically subtle and the result is not well-defined
unless appropriate regularity conditions are imposed.

Let us spell these out. The test fields $f^i$ must, as usual, be at
least continuous.  Conditions on the surface $S$ are less transparent.
First, the action (\ref{10}) may not be well-defined if the surface
$S$ and the graph $\g$ have an infinite number of intersections. This
is easily avoided by demanding that $S$ should also be analytic.  For,
each of our graphs $\g$ has only a finite number of analytic edges and
an analytic curve either intersects an analytic surface only at a
finite number of points or it is tangential to it.  Thus, analyticity
of $S$ will ensure that the sum on the right side of (\ref{11}) has
only a finite number of terms. Even then, however, ambiguities arise
if the surface $S$ has a boundary and an edge of the graph intersects
$S$ at one of the points on the boundary. Therefore, we are led to
require that $S$ does not have a boundary. (These regularity
conditions were also necessary in \cite{1} to ensure that the
two-dimensionally smeared triad {\it operator} be well-defined in the
quantum theory.)  The last condition arises because the fields
$(A_a^i, {E}^a_i)$ spanning the classical phase space are all
smooth. Because of this, if a surface $S_2$ is obtained from $S_1$
simply by deleting a set of measure zero, they define the same
functions (\ref{e2}) on the phase space. However, since the right side
of (\ref{10}) involves distributions, the value one obtains for $S_1$
may be different from that for $S_2$; some edges of $\g$ may intersect
$S_1$ precisely at those points which are missing in $S_2$. Therefore,
the limiting procedure can lead to well defined Poisson brackets only
if we remove this ambiguity.  We will do so by restricting the
permissible surfaces: we will only consider surfaces $S$ of the type
$S = {\bar S} - \partial {\bar{S}}$, where $\bar{S}$ is any compact,
analytic, 2-dimensional sub-manifold of $\S$ possibly with
boundary. This condition ensures that $S$ has no ``missing points in
its interior''.

With these technicalities out of the way, let us now associate with
each analytic surface $S$ in $\S$ of the above type and each test field
$f^i$ thereon a momentum variable $\Etwo$ on the phase space via
(\ref{e2}).  Following (\ref{10}), let us set the Poisson bracket
between these momenta and the configuration functions in Cyl to be:
\be\label{11}
\{ C_\g, \, {}^2\!E[S,f] \} 
= \frac{G}{2} \sum_{p}\, \sum_{I_p}\, \kappa({I_p})\, f^i(p)\, 
X^i_{I_p}\cdot c
\ee
where $X^i_{I_P}\cdot c$ is the result of the action of the $i$-th
left (resp. right) invariant vector field on the $I_p$-th copy of the
group if the $I_p$-th edge is pointing away from (resp. towards) the
surface $S$.  Note the structure of the right hand side. The result is
non-zero only if the graph $\g$ used in the definition of the
configuration variable $C_\g$ intersects the surface $S$ used to smear
the triad. If the two intersect, the contributions arise from the
action of right/left invariant vector fields on the arguments of $c$
associated with the edges at the intersection.

It only remains to specify the Poisson brackets between the new
momentum variables, $\Etwo$. In view of (\ref{2}), the obvious choice,
assumed implicitly in some of the early literature, is to set:
\be \label{12}
\{ {}^2\!{E}[S,f], \,\,  {}^2\!{E}[S',f']\} = 0. 
\ee
With this assumption, the Poisson brackets between all our fundamental
configuration and momentum variables are specified. If these brackets
satisfy the Jacobi identity, the resulting Lie algebra would offer a
point of departure for the quantum theory.

\subsection{Problem with the naive Poisson algebra}
\label{sec2.3}

However, as we will see in this sub-section, the brackets (\ref{11})
and (\ref{12}) are in fact inconsistent with the Jacobi identity.
Hence, one must develop a new quantization strategy. This task will be
carried out in section \ref{sec3}.

For simplicity, we will illustrate the problem with specific Poisson
brackets. Fix a two-surface $S$, two smearing fields, $f^i$ and $g^i$
thereon, and a closed loop $\a$ which has a finite segment tangential
to $S$ as in figure 1. The Poisson brackets of interest to us are
those between the momentum variables $\Etwo$ and $\Etwog$ and the
configuration variable $C_\a := T_\a$. (Here, $T_\a (A) =
\textstyle{1\over 2}\tr\, h_\a$, where the trace of holonomy is taken in
the
fundamental representation.) For concreteness, we parametrize and
orient the loop $\a$ as in the figure. We will present the calculation
in some detail because the final result is quite surprising.

\bigskip\bigskip

\centerline{
\hfill\vbox{\epsfxsize=4in\epsfysize=2.2in\epsfbox{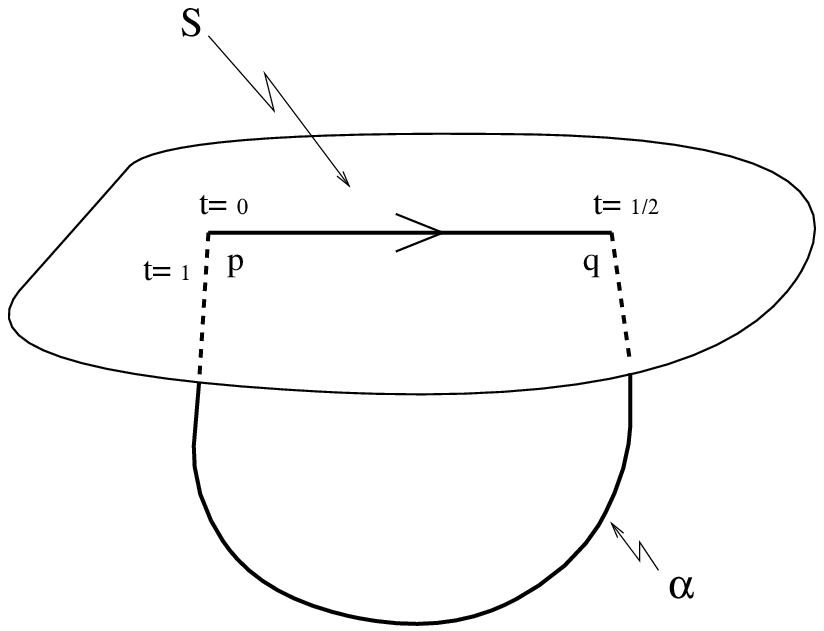}}\hfill}
\bigskip

\noindent
{\small {\bf Fig. 1} The loop $\alpha$ has two analytic segments, one
which lies in the surface $S$ and the other which lies entirely below
$S$. For definiteness, the loop $\alpha$ has been so parameterized that
the parameter $t$ runs from $0$ to $\frac{1}{2}$ along the segment in
$S$
and from $\frac{1}{2}$ to $1$ for the other segment. The surface is 
oriented so that the  segment $t \in [1/2,1]$ lies `below' $S$.}
\bigskip\bigskip

Let us begin by computing the Poisson bracket $\{\Ta,\Etwo\}$ which is
used repeatedly in the rest of the calculation. We have:
\ba
\{\Ta,\, \Etwo\}
&=&G\int \d^3\!x {f}_a^i(\x)\, \frac{\delta \Ta}{\delta A^i_a(\x)}\nl 
&=&G\int_S \d S^{ab}\teta_{abc}\, f^i(\s)
\int^1_0 \d t \,\dot{\a}^c(t)\,\delta^3(\s,{\vec \a}(t))\, 
\tr[h_\a(1,t)\tau^i h_\a (t,0)]\, , 
\ea
where $\s$ denotes a generic point in $S$. 

Therefore, for $\Ta$ and $S$ as in figure 1, we have 
\ba
\{\Ta,\, \Etwo\}
&=&G\int_S \d S^{ab}\,\teta_{abc}f^i\, \left[\int_0^{1/2}\d t\;
\dot{\a}^c(t)
\delta^3 (\s, {\vec \a}(t))\, {\tr} [h_\a(1,t) \tau^i h_\a(t,0)]
\right.\nl
&{}&\left.+ \int_{1/2}^1 \d t \; \dot{\a}^c(s)\delta^3(\s, \vec{\a}(t))
{\tr} [h_\a(1,t)\tau^i h_\a(t,0)]\right]\nl
&=&-\frac{G}{2} f^i(q){\tr}[h_\a (1/2,1/2)\tau^i]+\frac{G}{2} f^i(p)
{\tr}[\tau^i h_\a(1,0)]
\ea
where we have used the standard convention, $\int_0^\infty \d x\,
\delta (x) = \frac{1}{2}$, which can be justified from general
consistency considerations (see, e.g. \cite{1}). Therefore we have, 
\ba
\{\{\Ta,\, \Etwo\},\,\Etwog\}&=&\frac{G^2}{4}\, f^i(q) g^j(q)\,
\tr[h_\a (1/2,1/2)\tau^i\tau^j]\nl 
&-&\frac{G^2}{4}\, f^i(q) g^j(p)\, \tr[\tau^j h_\a
(1,1/2)\tau^ih_\a(1/2,0)]\nl
&+& \frac{G^2}{4}\, f^i(p)g^j(p)\, \tr[\tau^j\tau^i h_\a (1,0)]\nl
&-&\frac{G^2}{4} f^i(p) g^j(q) \tr[\tau^i h_\a
(1,1/2)\tau^jh_\a(1/2,0)]\, . 
\ea 

We are now ready to evaluate the left hand side of the Jacobi identity.
We have:
\ba {\rm J}
&:=&\{\{\Etwo,\, {}^2E[S,g]\},\, \Ta \}+\{\{\Ta,\,
\Etwo\},\,{}^2E[S,g]\} 
+ \{\{{}^2E[S,g],\, \Ta\},\,\Etwo\}\nl
&=&\{\{\Etwo,\, {}^2E[S,g]\},\, \Ta \}+
\frac{G^2}{4}\, f^i(q) g^j(q)\, \tr[h_\a (1/2,1/2)(\tau^i\tau^j-\tau^j
\tau^i)]\nl
&{}&+\frac{G^2}{4}\, f^i(p) g^j(p)\, \tr[(\tau^j\tau^i-\tau^i\tau^j)
h(1,0)]\nl 
&=&\{\{\Etwo,\, {}^2E[S,g]\}, \Ta \}+\frac{G^2}{4} f^i(q)
g^j(q) \e^{ijk}\,\tr[h_\a (1/2,1/2)\tau^k]\nl
&{}&+\frac{G^2}{4} f^i(p) g^j(p)\e^{ijk}\tr[\tau^k h_\a (1,0)] 
\ea
By inspection, the sum of the last two terms is generically non-zero.
Hence, the Jacobi identity will be violated if we {\it demand} that
the Poisson bracket $\{\Etwo,\, \Etwog\}$ between momenta must
vanish. Thus, the bracket defined by (\ref{11}) and (\ref{12}) fails
to be a Lie bracket and can not serve as the starting point for
quantization.

Let us summarize. The kinematical symmetries of the theory suggest
that we use as configuration variables elements of Cyl, based on
holonomies along curves, rather than the three-dimensionally smeared
functions ${}^3\!A[v]$ of section \ref{sec2.1}. This is a viable strategy
because a (background-independent) functional calculus is readily
available on Cyl.  It is then natural to use two-dimensionally smeared
triads, $\Etwo$ as momenta because their Hamiltonian vector fields map
Cyl to itself. Our task then is to define a consistent kinematical
framework using these variables.  Since the action of the Hamiltonian
vector fields generated by $\Etwo$ is well-defined on Cyl, the
Lie-bracket between our configuration and momentum variables is
unambiguous; it is given by (\ref{11}). However, if we now require
--as seems natural at first-- that the momentum variables should
Poisson-commute, the Jacobi identity is violated. Thus, we do not have
a Lie algebra which can serve as the point of departure for quantum
theory. Hence a new strategy is needed.

{\it Remark}: In the early literature on the subject, manifest
$SU(2)$-gauge invariance was often at the forefront. The momentum
functions were then taken to be traces of products of $e_{ab}^i$ and
holonomies (the $T^0_\alpha$ and the $T^a_\alpha$ variables of Rovelli
and Smolin \cite{11}) or smeared versions thereof (the so-called
`strip variables', associated with foliated, 2-dimensional strips
(see, e.g., \cite{alm2t})). They are again linear in the triads
$e_{ab}^i$ but also depend on the connections $A_a^i$ to ensure gauge
invariance. A careful examination shows that the analog of problem
with the Jacobi identity we just discussed exists also in that
setting. Thus, the problem is {\it not} an artifact of our use of
non-gauge invariant variables.

\section{Solution: A Consistent Lie algebra}
\label{sec3}

An algebraic approach a la Dirac is best suited to quantization
especially because we wish to use a restricted class, $\cyl$, of
smooth functions on the classical phase space as our configuration
variables.  We can begin by associating configuration operators
with elements of $\cyl$ and momentum operators with $\Etwo$.  To
construct the quantum algebra, however, we need to specify the
commutators between these operators. In the final theory, the
configuration operators always act by multiplication and must
therefore commute among themselves. The commutator of the momentum and
configuration operators are also unambiguous; they are dictated by the
bracket (\ref{11}).  Thus, what is needed is the commutator of
the momentum operators among themselves.

In section \ref{sec3.1}, we will see that one can extract this
information by exploiting the fact that our phase space has a
cotangent bundle structure. That is, one can define brackets between
the classical momentum variables $\Etwo$ such that a {\it true}
Lie-algebra results. However, it turns out that not all elements of
this Lie-algebra can be represented as {\it functions} on the phase
space.  Nonetheless, we will see that the Lie algebra does serve as a
viable point of departure for quantization. Furthermore, the resulting
quantum algebra is precisely the one used in Refs [1-14]. Thus there
is no anomaly in quantization; the {\it classical} brackets between
momentum variables already fail to vanish. The origin of this
non-commutativity is discussed in section \ref{sec3.2}.

\subsection{Quantization Strategy: Exploiting the cotangent bundle 
structure}
\label{sec3.1}

As we saw in section \ref{sec2.1}, our phase space is a cotangent
bundle over the configuration space $\A$ of connections. It is
therefore natural to try to repeat the strategy one uses for
quantization of simple systems where the classical phase space $\G$ is
a cotangent bundle over a finite dimensional manifold, say $\C$.

Let us first review that situation briefly (see, e.g., \cite{21,22}).
For such systems, one begins with the space $\F$ of suitably regular
functions $f$ on $\C$. Elements of $\F$ can be lifted to $\G$ to yield
phase space functions which are independent of momenta. These are the
configuration variables. The momentum variables $M(q,p)$ are functions
on $\G$ which are linear in momenta (i.e. in the fibers of $\G$):
$M_V(q,p) = V^ap_a$ for {\it some} vector field $V^a$ on $\C$. Thus,
there is a natural isomorphism between the space of momentum variables
and the space $\V$ of suitably regular vector fields on $\C$. The
Poisson brackets between these elementary phase space functions (which
yield the commutators between the elementary quantum operators) are
given by:
\ba \label{pb}
\{ f,\, f'\} = 0 &;&\quad \{ f,\, M_V \} = V\cdot f\nl
\{ M_V\, M_{V'}\}&=& M_{[V',V]}\, ,
\ea
where $V\cdot f$ is the action of the vector field $V$ on the function
$f$ and $[V,\, V']$ is the commutator of the two vector fields. Note
that these operations refer only to the structure of the configuration
space $\C$ rather than the phase space $\G$. They mirror the natural
Lie algebra structure that exists on pairs $(f, V)$ of functions and
vector fields on the configuration space $\C$:
\be \label{lb}
[(f,V), \, (f', V')] = (V'\cdot f - V\cdot f', \, [V',V])
\ee
Thus, in effect, in the quantum theory one associates configuration
operators with elements of $\F$ and momentum operators with elements
of $\V$ and the commutators between these operators are dictated by
the natural Lie bracket (\ref{lb}) which refers only to $\C$. This
description is completely equivalent to the one in terms of Poisson
brackets (\ref{pb}) but does not directly refer to operations on the
phase space $\G$.

Let us now return to the problem at hand. Now the space $\A$ of
(suitably regular) connections plays the role of $\C$ and the space
Cyl of cylindrical functions plays the role of $\F$. Thus, to complete
the Lie algebra, we need to isolate the analog of $\V$, the space
suitable vector fields on $\A$. As the above discussion suggests, this
task can be completed by examining the momentum variables $\Etwo$.
Indeed, it follows from (\ref{11}) that, given the ring Cyl of
cylindrical functions, we can naturally associate a vector field
$X_{S,f}$ with the momentum variable $\Etwo$:
\be \label{X}
\Etwo\, \mapsto \, X_{S,f}\quad ; \quad X_{S,f}\cdot C_\g = 
\frac{1}{2} \sum_{p}\, \sum_{I_p}\, \kappa({I_p})\, f^i(p)\, 
X^i_{I_p}\cdot c
\ee
for any cylindrical function $C_\g$ based on a graph $\g$. However,
since $\A$ is infinite dimensional, it is important to specify the
sense in which $X_{S,f}$ is a vector field: $X_{S,f}$ {\it is a
derivation on the ring of cylindrical functions}. That is,
$$ X_{S,f}: {\rm Cyl} \, \longrightarrow \, {\rm Cyl} $$
such that the map is linear and satisfies the Leibnitz rule:
\ba
X_{S,f} \cdot (C_\g + \lambda C^\prime_\g) &=& X_{S,f}\cdot C_\g +
\lambda X_{S,f} \cdot C^\prime_\g \nl 
X_{S,f}\cdot (C_\g C^\prime_\g) &=&
C_\g\, X_{S,f}\cdot C^\prime_\g + (X_{S,f}\cdot C_\g)\, C'_\g
\ea
for all cylindrical functions%
\footnote{Note that there is no loss of generality in assuming that 
two cylindrical functions are cylindrical with respect to the same
graph.  Given $C_\g$ and $C'_{\g'}$ where $\g$ and $\g'$ are distinct
graphs, one can just consider a larger graph $\g''$ which contains all
the vertices and edges of the two graphs. Then, the two given
functions are cylindrical with respect to $\g''$.}
$C_\g$ and $C'_\g$ and complex numbers $\lambda$. Finally, note that
the commutator between two derivations on Cyl is well-defined and is
again a derivation on Cyl. Furthermore, these derivations form a
Lie-algebra; the commutator bracket automatically satisfies the Jacobi
identity.

Hence, it is natural to use for $\V$ the vector space of derivations
on Cyl generated by $X_{S,f}$. That is, $\V$ will be the vector space
consisting derivations on Cyl resulting from finite linear
combinations and a finite number of commutators of $X_{S,f}$. With
this choice of $\V$ and with $\F = \cyl$, the analog
\be\label{lb2}
\left[(C_\g, X_{S,f}), \, (C'_{\g'}, X_{S', f'})\right] = 
\left( X_{S',f'}\cdot 
C_\g - X_{S, f}\cdot C'_{\g'},\,\, [X_{S',f'},\, X_{S,f}\,]\right)
\ee
of (\ref{lb}) is a Lie-bracket for all cylindrical functions $C_\g$
and $C_{\g'}$ and vector fields $X_{S,f}$ and $X_{S',f'}$ in
$\V$. This is the Lie-algebra we were seeking.  To go over to the
quantum theory, with each element of $\cyl$, we can associate a
configuration operator and with each element of $\V$, a momentum
operator. The commutators between these operators can be taken to be
$i\hbar$ times the classical Lie bracket (\ref{lb2}).  Furthermore, it
is transparent from paper I that this quantum algebra is faithfully
represented by operators on $\H = L^2(\Ab ,\d\mu_\circ)$.  Thus, the
quantum theory of refs [1-14] in fact results from `quantization of
this classical Lie-algebra'. In particular, there is no anomaly
associated with this quantization.

Let us summarize. For simple finite dimensional systems, there are two
equivalent routes to quantization, one starting from the Poisson
algebra of configuration and momentum functions on the phase space and
the other from functions and vector fields on the configuration
space. It is the second that carries over directly to the present
approach to quantum gravity. 

However, there are some important differences between the situation in
the present case and that in simple finite dimensional examples. We
will conclude this sub-section with two remarks on these differences.

i) In finite dimensional examples, $\V$ is generally taken to be the
space of {\it all} smooth vector fields on the configuration
space. Here, on the other hand, we only considered those derivations
which can be generated from the basic vector fields $X_{S,f}$ by
taking their finite linear combinations and a finite number of Lie
brackets. What motivated this restriction?  Could we have allowed {\it
all} derivations on $\cyl$ and still obtained a Lie algebra? The
answer is in the affirmative. However, that procedure would have been
inconvenient for two reasons.

First, whereas the $X_{S,f}$ are in one to one correspondence with the
momentum functions $\Etwo$ of (\ref{e2}) on the phase space, as we
will see below, a generic derivation need not correspond to {\it any}
phase space function. In the quantization procedure, on the other
hand, it is convenient --and, for analyzing the classical limit, even
essential-- to have a correspondence between `elementary operators'
with which one begins and phase space functions \cite{21,22}.  Second, if
one allows {\it all} derivations, one must specify relations between
them which are to carry over to the quantum theory. Indeed, even in
finite dimensional systems, such relations exist and give rise to
certain anti-commutation relations which ensure that the operators
corresponding to functions $f$ and vector fields $V$ and $fV$ are
correctly related \cite{21,22}. In the present case, the task of
specifying all such relations would have been formidable.  By starting
with a `small' class of vector fields $X_{S,f}$, we avoid both sets of
difficulties in one stroke. (The strategy we chose is, in some ways,
analogous to the text-book treatment of quantization of a particle in
the Euclidean space, where one builds the Lie-algebra from just
constant vector fields $\partial/\partial x$, $\partial/\partial y$,
and $\partial/\partial z$.)
 
ii) In the finite dimensional case, there is a one to one
correspondence between suitably regular vector field $V^a$ and
momentum functions $V^ap_a$ on the phase space. In the present case,
this correspondence continues to hold for the basic vector fields
$X_{S,f}$ which generate $\V$. However, it does not extend to general
elements of $\V$. A simple example is provided by the commutator
$[X_{S,f},\,X_{S',f'}]$ where $S$ and $S'$ intersect on a one
dimensional line. It follows immediately from (\ref{X}) that the
commutator is again a derivation on $\cyl$ but its action is
non-trivial {\it only} on graphs with edges passing through the
intersection of $S$ and $S'$. That is, while the vector fields
$X_{S,f}$ and $X_{S',f'}$ each have a two-dimensional support, the
commutator has only one dimensional support. In particular, therefore,
it can not be expressed as finite a linear combination of our basic
vector fields and does not define a linear combination of momentum
functions $\Etwo$.

One might imagine going around this difficulty by extending the
definition of momentum functions.  For instance, one might associate
functions $\Etwo$ and vector fields $X_{S,f}$ not only with analytic
surfaces $S$ but also piecewise analytic ones. This strategy brings
with it additional complications because the $\kappa(I_p)$ in
(\ref{10}) are now ambiguous for edges $e_I$ passing through the
``corners'' at which the piecewise analytic $S$ fails to be
analytic. In simple situations, one can give a recipe to remove these
ambiguities. Then, the commutator can be expressed as a linear
combination of vector fields $X_{S^{(J)}, f^{(J)}}$ associated with
piecewise analytic surfaces $S^{(J)}$ constructed from $S$ and
$S'$. However, because the intersection of $S$ and $S'$ is one
dimensional and fields ${E}^a_i$ constituting the phase space are all
smooth, the corresponding linear combination of momentum functions
${}^2\!E[S^{(J)}, f^{(J)}]$ simply vanishes! Thus, even after
extending the definitions of momentum functions and basic vector
fields, one can not establish a one to one correspondence between the
vector fields and momentum functions. Therefore, we have refrained
from carrying out this extension.

Note that, in spite of these differences, the quantization strategy is
sound. The classical Lie algebra --which leads to the algebra of
quantum operators-- is generated by functions $C_\g$ on $\A$ and
vector fields $X_{S,f}$ on $\A$. Each of these generators defines, in
a one to one fashion, functions $f(A,E):= C_\g(A)$ and $M(A,E) := \int_S
e_{abi}f^i \d S^{ab}$ on the phase space. These functions are complete,
i.e., suffice to separate the points of the phase space. To promote a
classical observable to a quantum operator, we can first express it in
terms of these basic functions and then carry the expression
over to the quantum theory. The procedure carries with it only the
standard factor ordering and regularization ambiguities. Thus,
because the set of generators of the classical Lie-algebra is
`sufficiently large', the fact that some derivations do not correspond
to phase space functions does not create an obstacle.

\subsection{Origin of non-commutativity} 
\label{sec3.2}

Since we now have the Lie algebra on which quantization can be based,
we can probe the origin of non-commutativity of the two-dimensionally
smeared triad operators defined in \cite{1}. At the classical
level, the key question is of course whether the vector fields
$X_{S,f}$ commute on $\cyl$. For definiteness, let us choose the
orientation of the edges of $\g$ in such a way that all the edges
intersecting $S$ are outgoing at the intersection points. Then the
action of $X_{S,f}$ on any cylindrical function $C_\g$ based on $\g$
is given by
\be
X_{S,f}\cdot C_\g =\frac{1}{2}\sum_p
f_i(p)\left(\sum_{I_p^u}X^i_{I_p^u} -
\sum_{I_p^d}X^i_{I_p^d}\right)\cdot c .
\ee
Here the superscript $u$ (`up') refers to edges which lie `above' $S$
and $d$ (`down') to the edges which lie below (since $\S$ and $S$ are
both oriented, this division of edges can be made unambiguously) and,
since the edges are all `outgoing', $X^i$ are the left-invariant
vector fields (on the corresponding copy of the group in the argument
of $c$.) It is straightforward to compute the commutator between
$X_{S,f}$ using the fact that the left invariant vector fields $X^i$
on $SU(2)$ satisfy $[X^i, X^j] = \epsilon^{ijk} X^k$:

\be
\left[X_{S,f},\,  X_{S^\prime,f^\prime}\right]\cdot C_\g =
\frac{1}{4}\sum_{\bar{p}} f^i(\bar{p}){f^\prime}^j(\bar{p})
\epsilon_{ijk}\left(\sum_{I_{\bar{p}}^{uu^\prime}}
X^k_{I_{\bar{p}}^{uu^\prime}}
-\sum_{I_{\bar{p}}^{ud^\prime}}X^k_{I_{\bar{p}}^{ud^\prime}}-
\sum_{I_{\bar{p}}^{du^\prime}}X^k_{I_{\bar{p}}^{du^\prime}}+
\sum_{I_{\bar{p}}^{dd^\prime}}X^k_{I_{\bar{p}}^{dd^\prime}}\right)
\cdot c
\label{COM}
\ee
where the sum extends {\it only} on vertices $\bar{p}$ that lie on the
{\it intersection} of the surfaces $S$ and $S^\prime$, and,
$I_{\bar{p}}^{ud^\prime}$, for example, denotes an edge passing though
point $\bar{p}$ that lie `above' $S$ and below $S^\prime$. Thus, if
$\g$ has edges that intersect $S$ and $S'$ on the one-dimensional
curve on which $S$ and $S'$ themselves intersect, the commutator will
fail to vanish in general. Such intersections are of course
non-generic. Nonetheless, the subspace of $\cyl$ on which is the
commutator has a non-trivial action is infinite dimensional.  This
non-commutativity is simply mirrored to the triad operators in
quantum theory; thus the structure found in \cite{1} is not surprising
from the classical perspective.

Nonetheless, since the non-commutativity between quantum Riemannian
structures is a striking feature, let us probe equation (\ref{COM})
further. 

Recall, first that the functions $\Etwo$ were obtained by a limiting
procedure from the three-dimensionally smeared functions
$\Ethree$. Since these $\Ethree$ are linear in the triad, they are
associated, in a one-to-one manner with the vector fields $\Xthree$ on
$\A$:
\be \Xthree \cdot g := 
\int_{\S} \d^3x f^i_a(\x) \frac{\delta g}{\delta{A_a^i(\x)}}.
\ee
Although, as we saw in section \ref{sec2.2}, the action of these
vector fields does not preserve $\cyl$, they do have a well-defined
action on $\cyl$ (and in fact preserves the space of all smooth
functions on $\A$).  Furthermore, it is clear from equation (\ref{10})
that our fundamental vector fields $X_{f,S}$ arise as limits of
$\Xthree$,
\be X_{S,f}\cdot C_\g = \lim_{\epsilon\mapsto 0}\, 
{}^3\!X[{}^\epsilon\!f] \cdot C_\g\, ,
\ee
for all cylindrical functions $C_\g$. Now, it is obvious from their
definition that the vector fields $\Xthree$ commute. So, it is at
first puzzling that the $X_{S,f}$ do not. How does this arise?
After all, we have
\be \lim_{\epsilon^\prime \mapsto 0} \lim_{\epsilon\mapsto 0}\,
[{}^3\!X[{}^{\epsilon^\prime}\!f^\prime], \, \, 
{}^3\!X[{}^\epsilon\!f]\, ]\cdot C_\g\, = 
\, \lim_{\epsilon\mapsto 0} \lim_{\epsilon^\prime
\mapsto 0}\, [{}^3\!X[{}^{\epsilon^\prime}\!f^\prime], \, \,
{}^3\!X[{}^\epsilon\!f]\, ]\cdot C_\g\, = 0 \ee
for all $C_\g$ since the commutators vanish before taking the limit.
Note however, that the commutator $[X_{S^\prime, f^\prime}, X_{S,f}]$
does not result from either of these limits. Rather, it is given by
\ba & &[X_{S^\prime,f^\prime}\, ,\, X_{S,f}] \cdot C_\g \nl
&=& \lim_{\epsilon^\prime \mapsto 0} 
{}^3\!X[{}^{\epsilon^\prime}\!f^\prime] \, 
\left(\lim_{\epsilon\mapsto 0} {}^3\!X[{}^\epsilon\!f]\, 
\cdot C_\g\right)\, - 
\, \lim_{\epsilon\mapsto 0} 
{}^3\!X[{}^\epsilon\!f] \,  \left(\lim_{\epsilon^\prime
\mapsto 0}{}^3\!X[{}^{\epsilon^\prime}\!f^\prime]\, \cdot C_\g\,\right)
\ea
By expanding out the two terms one can see explicitly that the
non-commutativity arises because, while acting on cylindrical
functions, the action of the vector fields ${}^3\!X$ does not commute
with the operation of taking limits. 

To summarize, the regularization procedure which enables us to pass to
$X_{S,f}$ from $\Xthree$ is quite subtle and gives rise to a striking
contrast between $\Xthree$ and their limits $X_{S,f}$. First, the
action of $\Xthree$ preserves the space of smooth functions on $\A$
but fails to preserve the sub-space $\cyl$ thereof. The action of
$X_{S,f}$, on the other hand, leaves $\cyl$ invariant but is not even
well-defined on more general smooth functions. Secondly, since
$\Ethree$ depend only on the triads and not on connections, the vector
fields $\Xthree$ are constant with respect to the affine structure of
$\A$. However, this correspondence does {\it not} hold once we bring
in singular smearing fields: even though $\Etwo$ depend only on the
triads ${E}^a_i$ (i.e., have no direct dependence on connections
$A_a^i$), the vector fields $X_{S,f}$ are {\it not} constant. In fact
they fail to commute and this non-commutativity is then directly
reflected in that of the smeared triad operators in the quantum
theory.

\section{Discussion}
\label{sec4}

Consider classical systems in which the phase space has a natural
cotangent bundle structure. To quantize such a system, one can proceed
in the following steps: i) Choose a preferred set of configuration
variables, i.e., functions on the configuration space $\C$; ii) Choose
a set of preferred vector fields $X$ on $\C$ which are closed under
the Lie bracket and whose action leaves the space of configuration
variables invariant. The space of these preferred functions and vector
fields is automatically endowed with a natural Lie bracket; iii)
Associate with each configuration variable in the chosen set a
configuration operator and with each vector field in the chosen set a
momentum operator, require that their commutators be $i\hbar$ times
the Lie brackets of their classical counterparts and construct the
abstract operator algebra they generate; and, iv) Find a
representation of this algebra.  (For a more complete description,
see, e.g., \cite{21,22}.) For this procedure to lead us to a useful
quantum theory, however, it is necessary that the space of functions
and vector fields considered be `sufficiently large'. For example, the
configuration variables together with the momentum functions defined
by the vector fields should suffice to separate the points of the
phase space. Only then would one have a reasonable chance of promoting
a sufficiently large class of classical observables to quantum
operators (modulo the usual factor ordering ambiguities.)

For systems with a finite number of degrees of freedom, it is
generally straightforward to implement these steps. For example, for
configuration variables one can choose smooth functions of compact
support on $\C$ and require that the chosen vector fields also be
smooth and of compact support. For systems with an infinite number of
degrees of freedom, on the other hand, the choices are not so
stream-lined. 

In non-perturbative quantum gravity one is guided by the invariances
of the theory. Let us begin by recalling the general setting. The
configuration space can be taken to be the space $\A$ of all smooth
$SU(2)$ connections $A_a^i$ on a three-manifold $\Sigma$. Cylindrical
functions --i.e. functions which depend on the connection only through
its holonomies along edges of finite graphs-- suffice to separate
points of $\A$. Since holonomies are in effect the `raison d'\^etre'
of connections, it is then natural to choose the space $\cyl$ of
cylindrical functions as the space of configuration variables. The
action of the vector fields should leave $\cyl$ invariant. Therefore,
they have to be chosen from among derivations on the ring $\cyl$. The
requirement that the space be `sufficiently large' leads us to
consider the derivations which arise from functions on the phase space
which are linear in momenta $e_{ab}^i$. Together, these considerations
led us to the momentum functions $\Etwo := \int_S e_{ab}^i f_i
\d S^{ab}$ obtained by smearing the triads $e_{ab}^i$ by test field
$f_i$ on {\it two} dimensional surfaces and the corresponding vector
fields $X_{S,f}$.  Thus, if we choose for vector fields the
derivations on $\cyl$ which are obtained by taking finite linear
combinations of commutators between $X_{S,f}$, we satisfy all the
requirements to obtain a Lie-algebra which can serve as the point of
departure for quantization.

This construction is natural in the sense that it does not involve any
background structures: the connection 1-forms are integrated on
one-dimensional curves, while the triad two-forms are integrated over
two-dimensional surfaces. However, since we are effectively smearing
fields in one or two dimensions, rather than three, our elementary
variables are `singular' in a certain sense: compared to the standard
procedure followed in Minkowskian field theories, we have let the
smearing fields themselves be distributional. Therefore, care is
needed in calculating Poisson brackets between these configuration and
momentum variables. In particular, as discussed in section \ref{sec3.2},
although the three-dimensionally smeared triads {\it do} Poisson
commute with one another, the $\Etwo$ do not. Indeed, as we saw in
section \ref{sec2.3}, if we simply assume commutativity, we are led to a
violation of the Jacobi identity! Thus, the lower dimensional
smearings do lead to features which are at first counter
intuitive. However, the procedure we followed is well-defined and
internally consistent and one can proceed with quantization along the
steps listed in the beginning of this section. The result is precisely
the quantum theory that was  developed in references
\cite{1,2,3,4,5,6,7,8,9,10,rs,b,hall,alm2t}.
 
For completeness, let us now sketch how this comes about.  Having
chosen the preferred class of configuration variables and vector
fields, it is straightforward to construct the algebra of quantum
operators. To find its representations, one can first focus on the
Abelian algebra of configuration operators. One can show that, in {\it
any} representation of this algebra, the Hilbert space is the space of
square-integrable functions (for some regular measure) on a certain
space, $\Ab$, which can be thought of as a `completion' of $\A$ in
an appropriate sense. The configuration operators act, as expected, by
multiplication. If we further require that the momentum operators act
simply by derivation, mirroring the action of the vector fields
$X_{S,f}$ on $\cyl$ in the classical theory, then the requirement that
the two sets of operators be self-adjoint leads us to the measure
$\mu_\circ$ on $\Ab$, referred to in the Introduction. In this quantum
theory, the commutators between the momentum operators simply mirror
the commutators between the vector fields $X_{S,f}$ in the classical
theory. Thus, there is no anomaly.

It is nonetheless striking that the smeared triad operators do not
necessarily commute. For, the triads are the fundamental fields from
which all Riemannian structures are built and their non-commutativity
implies that geometrical quantities such as the area operators also
fail to commute. Hence, quantum Riemannian geometry is now
intrinsically non-commutative. As emphasized in the Introduction, this
implies, in particular, that in this approach, (at least the naive)
metric representation does not exist.  What is the central assumption
responsible for this surprising feature?  It is that the configuration
variables be cylindrical functions, or, in the gauge invariant
context, traces of holonomies (i.e., Wilson loops). The assumption
seems rather tame at first. Furthermore, as indicated in the
Introduction, it is strongly motivated by the invariances of the
theory. Yet, once it is made, a series of natural steps lead us to
non-commutativity of quantum Riemannian structures.  These 
steps do involve additional assumptions but these appear to be
relatively minor, and of a rather technical nature; if desired, one
could weaken or justify them. Thus, it appears that the surprising
features of quantum geometry --non-commutativity, the polymer-like,
one-dimensional nature of fundamental excitations and the discreteness
of spectra of geometric operators-- can in essence be traced back to
to the assumption on the configuration variables and to the gauge and 
diffeomorphism invariance of the theory.

In the classical and quantum theory discussed in this series of papers
we have focussed on the kinematical structures. Consider, for example,
the area of a fixed surface $S$. In the classical theory, it is a
function on the full --rather than the reduced-- phase space and, in
the quantum theory, an operator on the kinematic --rather than the
physical-- Hilbert space. It does not commute with the constraints and
is thus not a `Dirac observable'. Therefore, the physical implications
of the non-commutativity are not immediately transparent. To clarify
this issue, let us re-examine the situation in classical general
relativity.  There, differential geometry provides us with a
mathematical formula to compute the area of {\it any} surface. To
relate it to physical measurements, we specify the surface
operationally, typically using matter fields. It is natural to expect
that the situation would be similar in the quantum theory. Given {\it
any} surface $S$, quantum Riemannian geometry provides us with an
operator $\hat{A}_S$. To relate it to observable quantities, we would
only have to specify the surface operationally. Therefore, the result
on non-commutativity of areas {\it should} have observable
consequences: the Heisenberg uncertainty principle prevents us from
measuring areas of intersecting surfaces with arbitrary accuracy.

Can one reconcile this with the fact that we have encountered no such
limit in the laboratory? The answer is in the
affirmative. Furthermore, the argument can be made at a sufficiently
general level, without recourse to a detailed measurement theory%
\footnote{Incidentally, such a theory will have interesting twists of its
own in the gravitational context since we can no longer use
`infinitely heavy' instruments that Bohr and Rosenfeld were forced to
introduce in their analysis of the Heisenberg uncertainties
associated with the quantum electro-magnetic field. So, a priori it is
not clear that even the area of a single surface can be measured with 
an arbitrary accuracy. (See, e.g. \cite{23}.) For our argument, however,
this subtlety is not relevant.}, 
which is fortunate since such a theory is yet to be developed in
quantum gravity. Consider two {\it macroscopic} surfaces $S_i$, i.e.,
two surfaces and a semi-classical state in which their areas are very
large compared to the Planck area. Then the expectation value of their
areas may be roughly estimated as $\langle A(S_i)\rangle\approx
l_P^2\,N(S_i)m_i$, where $m_i$ is an `average' spin, $N(S_i)$ is the
`effective' number of transverse intersections between the graph
underlying the semi-classical state and the surface $S_i$, and
$l_P^2:=G\hbar$ is the Planck area.  In the same way, the expectation
value of the commutator of the area operators is approximately
$l_P^4\,N(S_1 \cap S_2)k$, where $k$ is the corresponding `effective'
spin. (In a semi-classical state, $m_i$ and $k$ are of the same order.) 
We can estimate the relative uncertainty as follows:
\be \label{uncertainty} 
\frac{\Delta (A(S_1))}{\langle A(S_1)\rangle} 
\frac{\Delta (A(S_2))}{\langle A(S_2)\rangle}\,\, \geq \,\,
 \frac{k\,N(S_1 \cap S_2)}{m_1m_2\,N(S_1) N(S_2)} 
\ee 
Now, the intersection of two surfaces is a subset of measure zero in
either one of the surfaces. Therefore, for a semi-classical state, $N(S_1
\cap S_2)$ must be negligible compared to $N(S_1)$ or $N(S_2)$, and the
inequality should be close to being saturated, whence
the relative uncertainty should also be negligible. Thus, because the 
commutator of two area operators is a distribution with only one-dimensional
support, the uncertainties associated with their non-commutativity are
completely negligible on semi-classical states.

Finally, note that the non-commutativity between geometric structures
we have encountered here is quite different from that of
non-commutative geometry of Connes and others. In our approach, the
manifold itself is in tact, the notions of curves $\alpha$, surfaces
$S$ and regions $R$ of the three-manifold are all well-defined. (If we
consider matter fields, we can even specify these objects
operationally.) Hence they serve as well-defined labels for the
length, area and volume operators, $\hat{\ell}_\alpha, \hat{A}_S,
\hat{V}_R$, respectively. Non-commutativity refers to these operators;
it refers to quantum Riemannian structures, where the term `quantum'
is used in the old-fashioned, text-book sense. In Connes' framework,
by contrast, the non-commutativity occurs at a mathematically deeper
level: `the manifold itself becomes non-commutative'. More precisely,
one begins with the observation that for standard manifolds, the
manifold structure is coded in the {\it Abelian} $C^\star$-algebra of
smooth functions and generalizes geometry by considering instead {\it
non-Abelian} $C^\star$ algebras (equipped with certain additional
structures). Note, however, that in this general framework, there is
no `quantization' in the traditional sense, no obvious place for
$\hbar$. Once the algebra is non-commutative, `points of the manifold
disappear' and there is no obvious meaning to curves $\alpha$,
surfaces $S$ or regions $R$ and hence to the associated observables,
$\hat{\ell}_\alpha, \hat{A}_S$ and $\hat{V}_R$. Thus, although in both
approaches algebraic methods are used heavily to introduce geometric
notions, the two meanings of `non-commutativity' are quite different from 
one another.

\section*{Acknowledgments}

We would like to thank participants of the Erwin Schr\"odinger
workshop on Mathematical Problems of Non-perturbative Quantum Gravity
and especially among them Jerzy Lewandowski for numerous discussions.
This work was supported in part by the NSF grants PHY95-14240, INT
9722514 and by the Eberly research fund of Penn State University. In
addition, AC and JAZ were supported in part by DGAPA of UNAM, and
AC by CONACyT (M\'exico) Ref. No. I25655-E.

\end{document}